\documentclass[12pt]{article}
\usepackage{graphicx}
\usepackage{lscape}
\usepackage{citesort}
\usepackage{amssymb}
\usepackage{appendix}
\usepackage{multirow}

\begin{document}
\def\qq{\langle \bar q q \rangle}
\def\uu{\langle \bar u u \rangle}
\def\dd{\langle \bar d d \rangle}
\def\sp{\langle \bar s s \rangle}
\def\GG{\langle g_s^2 G^2 \rangle}
\def\Tr{\mbox{Tr}}
\def\figt#1#2#3{
        \begin{figure}
        $\left. \right.$
        \vspace*{-2cm}
        \begin{center}
        \includegraphics[width=10cm]{#1}
        \end{center}
        \vspace*{-0.2cm}
        \caption{#3}
        \label{#2}
        \end{figure}
    }

\def\figb#1#2#3{
        \begin{figure}
        $\left. \right.$
        \vspace*{-1cm}
        \begin{center}
        \includegraphics[width=10cm]{#1}
        \end{center}
        \vspace*{-0.2cm}
        \caption{#3}
        \label{#2}
        \end{figure}
                }

\def\ds{\displaystyle}
\def\beq{\begin{equation}}
\def\eeq{\end{equation}}
\def\bea{\begin{eqnarray}}
\def\eea{\end{eqnarray}}
\def\beeq{\begin{eqnarray}}
\def\eeeq{\end{eqnarray}}
\def\ve{\vert}
\def\vel{\left|}
\def\ver{\right|}
\def\nnb{\nonumber}
\def\ga{\left(}
\def\dr{\right)}
\def\aga{\left\{}
\def\adr{\right\}}
\def\lla{\left<}
\def\rra{\right>}
\def\rar{\rightarrow}
\def\lrar{\leftrightarrow}
\def\nnb{\nonumber}
\def\la{\langle}
\def\ra{\rangle}
\def\ba{\begin{array}}
\def\ea{\end{array}}
\def\tr{\mbox{Tr}}
\def\ssp{{\Sigma^{*+}}}
\def\sso{{\Sigma^{*0}}}
\def\ssm{{\Sigma^{*-}}}
\def\xis0{{\Xi^{*0}}}
\def\xism{{\Xi^{*-}}}
\def\qs{\la \bar s s \ra}
\def\qu{\la \bar u u \ra}
\def\qd{\la \bar d d \ra}
\def\qq{\la \bar q q \ra}
\def\gGgG{\la g^2 G^2 \ra}
\def\q{\gamma_5 \not\!q}
\def\x{\gamma_5 \not\!x}
\def\g5{\gamma_5}
\def\sb{S_Q^{cf}}
\def\sd{S_d^{be}}
\def\su{S_u^{ad}}
\def\sbp{{S}_Q^{'cf}}
\def\sdp{{S}_d^{'be}}
\def\sup{{S}_u^{'ad}}
\def\ssp{{S}_s^{'??}}

\def\sig{\sigma_{\mu \nu} \gamma_5 p^\mu q^\nu}
\def\fo{f_0(\frac{s_0}{M^2})}
\def\ffi{f_1(\frac{s_0}{M^2})}
\def\fii{f_2(\frac{s_0}{M^2})}
\def\O{{\cal O}}
\def\sl{{\Sigma^0 \Lambda}}
\def\es{\!\!\! &=& \!\!\!}
\def\ap{\!\!\! &\approx& \!\!\!}
\def\md{\!\!\!\! &\mid& \!\!\!\!}
\def\ar{&+& \!\!\!}
\def\ek{&-& \!\!\!}
\def\kek{\!\!\!&-& \!\!\!}
\def\cp{&\times& \!\!\!}
\def\se{\!\!\! &\simeq& \!\!\!}
\def\eqv{&\equiv& \!\!\!}
\def\kpm{&\pm& \!\!\!}
\def\kmp{&\mp& \!\!\!}
\def\mcdot{\!\cdot\!}
\def\erar{&\rightarrow&}


\def\simlt{\stackrel{<}{{}_\sim}}
\def\simgt{\stackrel{>}{{}_\sim}}


\title{
         {\Large
                 {\bf
                     Analysis of the  $\Lambda_{b}\rar
                     \Lambda \ell^+\ell^- $ decay in QCD
                 }
         }
      }

\author{\vspace{1cm}\\
{\small T. M. Aliev$^a$ \thanks {e-mail:
taliev@metu.edu.tr}~\footnote{permanent address:Institute of
Physics,Baku,Azerbaijan}\,\,, K. Azizi$^b$ \thanks {e-mail:
kazizi@dogus.edu.tr}\,\,, M. Savc{\i}$^a$ \thanks
{e-mail: savci@metu.edu.tr}} \\
{\small $^a$ Physics Department, Middle East Technical University,
06531 Ankara, Turkey} \\
{\small $^b$ Physics Division, Faculty of Arts and
Sciences, Do\u gu\c s University} \\
{\small Ac{\i}badem-Kad{\i}k\"oy, 34722 Istanbul} }
\date{}

\begin{titlepage}
\maketitle
\thispagestyle{empty}

\begin{abstract}
Taking into account the $\Lambda$ baryon distribution amplitudes and the most
general form of the interpolating current of the  $\Lambda_{b}$,
the semileptonic $\Lambda_{b}\rar \Lambda \ell^+\ell^- $
transition is  investigated in the framework of the light cone QCD sum
rules. Sum rules for all twelve form factors responsible for the $\Lambda_{b}\rar
\Lambda \ell^+\ell^- $ decay are constructed. The obtained results
for the form factors are used to compute the branching fraction.
A comparison of the obtained results with the existing predictions
of the heavy quark effective theory is presented.
The results of the branching ratio shows the detectability of this
channel at the LHCb in the near future is quite high.
\end{abstract}

~~~PACS number(s): 11.55.Hx, 13.30.-a, 14.20.Mr
\end{titlepage}

\section{Introduction}
Experimentally, the detection and isolation of the heavy baryons is simple
comparing to the light systems since having the heavy quark makes their beam
narrow. In the recent years, considerable experimental progress has been made
in the identification and spectroscopy of the heavy baryons containing a
heavy bottom or charm quark
\cite{1Mattson,2Aaltonen1,3Chistov,4Ocherashvili,5Abazov1,6Acosta,7Aubert1,8Solovieva}.
These evidences can be considered as a good
signal to search also the decay channels of the heavy baryons like,
$\Lambda_{b}\rar \Lambda \ell^+\ell^- $ at LHCb.
This rare channel,  induced by the flavor changing neutral currents (FCNC) of
$b\rightarrow s$ transition,
serves testing ground for the standard model  at loop level and is very sensitive to
the new physics  effects \cite{9aliev1}, such as supersymmetric
particles \cite{10susy}, light dark matter \cite{11darkmatter} and also
fourth generation of the quarks and extra dimensions, etc.
Moreover, this channel  can be inspected as a useful tool
in exact  determination of    the
Cabibbo-Kobayashi-Maskawa (CKM) matrix elements, $V_{tb}$ and
$V_{ts}$,   CP and T violations,
polarization asymmetries.

Theoretically, there are some works devoted to the analysis of the
heavy baryon decays, where practically in all of them the predictions   of the  heavy quark effective theory
(HQET) for form factors have been used. Transition form factors of the $\Lambda_{b} \rightarrow
\Lambda_{c}$ and $\Lambda_{c} \rightarrow \Lambda$ decays have been
studied in three points QCD sum rules in \cite{12Marques}, the
$\Lambda_{b}\rightarrow pl\nu$ transition form factors have also
been calculated via three point QCD sum rules in the context of the
heavy quark effective theory (HQET)  in \cite{13Huang} and in the
framework of the SU(3) symmetry and HQET in \cite{14Datta}. In the
present work, using the most general form of the interpolating
current for the $\Lambda_b$ and also the distribution amplitudes of
$\Lambda$ baryon, all form factors related to the electroweak
penguin and weak box diagrams describing the $\Lambda_{b}\rar
\Lambda \ell^+\ell^- $ are calculated in the frame work of the light
cone QCD sum rules in full theory. The obtained results for the form
factors are used to estimate the decay rate and branching ratio.
Regard that this transition has been investigated in \cite{15ming1}
and \cite{16ming2} also in the context of the HQET but the same
frame work using the distribution amplitudes of the $\Lambda$ and
$\Lambda_b$, respectively. Moreover, form factors, branching ratio
and di--lepton forward--backward asymmetries are studied in
\cite{yenison,17chen1,18chen2} also within the context of the HQET. In
\cite{19kmayprd,20kmyh,21kmz}, $\Sigma_{b,c}$ and $\Lambda_{b,c}$ to
nucleon transitions are also evaluated using the nucleon wave
functions in light cone QCD sum rules approach.

The plan of the paper is as follows: in section II,
the light cone QCD sum rules for  the form factors are obtained using the $\Lambda $
DA's. The  HQET relations among all form factors  are also discussed  in this section.
Section III is dedicated to the numerical analysis of the sum rules for the form factors as well as numerical
results of the decay rate and branching ratio.

\section{ Theoretical Framework}
The $\Lambda_{b}\rar \Lambda \ell^+\ell^- $ channel proceeds via FCNC
 $b\rightarrow s$ transition at quark level. The  effective Hamiltonian
describing the electroweak penguin and weak box diagrams related to this transition
can be written as :
\bea \label{ham}
{\cal H}_{eff} \es \frac{G_F~\alpha_{em} V_{tb}~V_{ts}^{^{*}}}{2\sqrt2~\pi}
\Bigg\{\vphantom{\int_0^{x_2}}  C_{9}^{eff}~ \bar{s}
\gamma_\mu (1-\gamma_5) b \bar l \gamma^\mu
 l +C_{10} ~\bar{s}
\gamma_\mu (1-\gamma_5) b \bar l \gamma^\mu \gamma_{5}l \nnb \\
\ek2 m_{b}~C_{7}\frac{1}{q^{2}} ~\bar{s} i \sigma_{\mu\nu}q^{\nu}
(1+\gamma_5) b \bar l \gamma^\mu l \Bigg\}~.  \eea To find the
amplitude, we need to sandwich this effective Hamiltonian between
the initial and final baryon states, i.e.,
$\langle \Lambda_b(p+q) \vert  {\cal H}_{eff}  \vert \Lambda(p)
\rangle $. From Eq. (\ref{ham}) we see that in calculation of the
$\Lambda_{b}\rar \Lambda \ell^+\ell^- $  decay amplitude, the
matrix elements,  $\langle \Lambda_b(p+q) \vert  \bar b \gamma_\mu
(1-\gamma_5) s\vert \Lambda(p) \rangle $ and  $\langle \Lambda_b(p+q)
\vert  \bar{b} i \sigma_{\mu\nu}q^{\nu} (1+ \gamma_5)
s\vert
\Lambda(p) \rangle $ are appeared. These matrix elements can be
parametrized  in terms of the twelve   form factors, $f_{i}$,
$g_{i}$, $f^T_{i}$ and $g^T_{i}$ in
the following manner:

\bea\label{matrixel1a}
\langle
\Lambda(p) \md  \bar s \gamma_\mu (1-\gamma_5) b \mid \Lambda_b(p+q)\rangle=
\bar {u}_\Lambda(p) \Big[\gamma_{\mu}f_{1}(Q^{2})+{i}
\sigma_{\mu\nu}q^{\nu}f_{2}(Q^{2}) + q^{\mu}f_{3}(Q^{2}) \nnb \\
\ek \gamma_{\mu}\gamma_5
g_{1}(Q^{2})-{i}\sigma_{\mu\nu}\gamma_5q^{\nu}g_{2}(Q^{2})
- q^{\mu}\gamma_5 g_{3}(Q^{2})
\vphantom{\int_0^{x_2}}\Big] u_{\Lambda_{b}}(p+q)~,
\eea
and
\bea\label{matrixel1b}
\langle \Lambda(p)\md \bar s i \sigma_{\mu\nu}q^{\nu} (1+ \gamma_5)
b \mid \Lambda_b(p+q)\rangle =\bar{u}_\Lambda(p)
\Big[\gamma_{\mu}f_{1}^{T}(Q^{2})+{i}\sigma_{\mu\nu}q^{\nu}f_{2}^{T}(Q^{2})+
q^{\mu}f_{3}^{T}(Q^{2}) \nnb \\
\ar \gamma_{\mu}\gamma_5
g_{1}^{T}(Q^{2})+{i}\sigma_{\mu\nu}\gamma_5q^{\nu}g_{2}^{T}(Q^{2})
+ q^{\mu}\gamma_5 g_{3}^{T}(Q^{2})
\vphantom{\int_0^{x_2}}\Big] u_{\Lambda_{b}}(p+q)~,
\eea
 where $Q^{2}=-q^{2}$.
 For calculation of these form factors we use the QCD sum rules approach. To obtain
the sum rules for the form factors in this approach, the following correlation
functions, the main  objects in this approach, are considered: \bea\label{T} \Pi^I_{\mu}(p,q) =
i\int d^{4}xe^{-iqx}\langle 0 \mid
T\{ J^{\Lambda_{b}}(0), \bar b(x) \gamma_\mu (1-\gamma_5) s(x))\}\mid  \Lambda(p)\rangle~,\nnb\\
\Pi^{II}_{\mu}(p,q) = i\int d^{4}xe^{-iqx}\langle 0 \mid
T\{ J^{\Lambda_{b}}(0),\bar{b}(x) i \sigma_{\mu\nu}q^{\nu} (1+ \gamma_5)
s(x)\}\mid
\Lambda(p)\rangle~,
\eea
where, $p$ represents the $\Lambda$'s momentum and $q$
is the transferred momentum and the  $J^{\Lambda_{b}}$ is interpolating current
of $\Lambda_{b}$. The most general form of the interpolating current of
$\Lambda_b$ baryon can be written as:
\bea\label{cur.N}
J^{\Lambda_{b}}(x)\es\frac{1}{\sqrt{6}}\epsilon_{abc}
\Bigg\{\vphantom{\int_0^{x_2}}2\Big[(q_{1}^{aT}(x)Cq_{2}^{b}(x))\gamma_{5}b^{c}(x)
+\beta(q_{1}^{aT}(x)C\gamma_{5}q_{2}^{b}(x))b^{c}(x)\Big] \nnb \\
\ar (q_{1}^{aT}(x)Cb^{b}(x))\gamma_{5}q_{2}^{c}(x)
+\vphantom{\int_0^{x_2}}
\beta(q_{1}^{aT}(x)C\gamma_{5}b^{b}(x))q_{2}^{c}(x) \nnb \\
\ar (b^{aT}(x)Cq_{2}^{b}(x))
\gamma_{5}q_{1}^{c}(x)
+\beta(b^{aT}(x)C\gamma_{5}q_{2}^{b}(x))q_{1}^{c}(x)\Bigg\}~,
\eea
where $q_{1}$ and $q_{2}$ are the $u$ and $d$ quarks, respectively,
$a,~b$ and $c$ are  color indexes and $C$ is the charge conjugation
operator.  The  $\beta$ is an arbitrary parameter with $\beta=-1$
corresponding to the Ioffe current.

In order to obtain the sum rules for the transition form factors,
we will calculate the aforementioned correlation functions in two
different ways, namely, physical (phenomenological) and theoretical
(QCD) sides and equate these two representations isolating the ground
state through the dispersion relation. Finally, to suppress the contribution
of the higher states and continuum, we will apply the Borel transformation
and continuum subtraction to both sides of the correlation function and
impose the quark hadron duality assumption.

The first step is to  calculate the  physical
side of the correlation functions. Saturating the correlation
functions with complete set of the intermediate states with the same quantum
numbers as the initial state, for the physical part of the correlation
function we obtain,
\bea\label{phys1}
\Pi_{\mu}^{I}(p,q)=\sum_{s}\frac{\langle 0\mid J^{\Lambda_{b}}(0)
\mid \Lambda_{b}(p+q,s)\rangle\langle
\Lambda_{b}(p+q,s)\mid  \bar b \gamma_\mu (1-\gamma_5) s \mid
\Lambda(p)\rangle}{m_{\Lambda_{b}}^{2}-(p+q)^{2}}+\cdots~,
\eea
\bea\label{phys1111}
\Pi_{\mu}^{II}(p,q)=\sum_{s}\frac{\langle 0\mid J^{\Lambda_{b}}(0)
\mid \Lambda_{b}(p+q,s)\rangle\langle
\Lambda_{b}(p+q,s)\mid \bar{b} i \sigma_{\mu\nu}q^{\nu} (1+ \gamma_5)
s \mid
\Lambda(p)\rangle}{m_{\Lambda_{b}}^{2}-(p+q)^{2}}+\cdots~,
\eea
where the dots $\cdots$ represent the contribution of the higher states and
continuum. The  vacuum  to the baryon matrix element of the
interpolating current,   $\langle0\mid
J^{\Lambda_{b}}(0)\mid \Lambda_{b}(p+q,s) \rangle$ is  written in terms of
the residue,   $\lambda_{\Lambda_{b}}$ as:
\bea\label{matrixel2}
\langle0\mid  J^{\Lambda_{b}}(0)\mid
\Lambda_{b}(p+q,s)\rangle=\lambda_{\Lambda_{Q}} \bar u_{\Lambda_{Q}}(p+q,s)~.
\eea
   Putting Eqs.
(\ref{matrixel1a}), (\ref{matrixel1b}) and (\ref{matrixel2})  in  Eqs.
(\ref{phys1}) and (\ref{phys1111}) and performing summation
over spins of the $\Lambda_{b}$ baryon using
\bea\label{spinor}
\sum_{s}u_{\Lambda_{b}}(p+q,s)\overline{u}_{\Lambda_{b}}(p+q,s)=\not\!p+\not\!q+m_{\Lambda_{b}}~,
\eea
 we get the following expressions for the correlation functions
\bea\label{phys2}
\Pi_{\mu}^{I}(p,q)\es
\lambda_{\Lambda_{b}} {\not\!p +\not\!q +m_{\Lambda_b}\over
m_{\Lambda_{b}}^{2}-(p+q)^{2}}
\Big\{\gamma_{\mu}f_{1} - i\sigma_{\mu\nu} q^{\nu} f_{2} +
q_{\mu} f_{3} \nnb \\
\ek \gamma_{\mu}\gamma_5 g_{1}
- i\sigma_{\mu\nu} q^{\nu} \gamma_5 g_{2}
+q_{\mu}\gamma_5 g_{3}
\vphantom{\int_0^{x_2}}\Big\} u_\Lambda(p) ~, \\ \nnb \\
\label{phys22}
\Pi_{\mu}^{II}(p,q)\es
\lambda_{\Lambda_{b}}{\not\!p +\not\!q +m_{\Lambda_b}\over
m_{\Lambda_{b}}^{2}-(p+q)^{2}}
\Big\{\gamma_{\mu}f_{1}^{T}- i \sigma_{\mu\nu}q^{\nu}f_{2}^{T} +
q^{\mu}f_{3}^{T}  \nnb \\
\ar \gamma_{\mu}\gamma_5
g_{1}^{T} + i \sigma_{\mu\nu}\gamma_5q^{\nu}g_{2}^{T}
- q^{\mu}\gamma_5 g_{3}^{T} \Big\} u_\Lambda(p)~.
\eea
Using the equation of motion and Eqs. (\ref{phys2}) and
(\ref{phys22}), we get the following final expressions
for the phenomenological sides of the correlation
functions:
\bea\label{sigmaaftera}
\Pi_{\mu}^{I}(p,q)\es
\frac{\lambda_{\Lambda_{b}}}{m_{\Lambda_{b}}^{2}-(p+q)^{2}}
\Big\{ 2 f_{1}(Q^{2})p_\mu +
2 f_{2}(Q^{2}) p_\mu\not\!q
+\Big[ f_2(Q^2) + f_3(Q^2)\Big] q_\mu\not\!q \nnb \\
\ek 2 g_1(Q^2) p_{\mu}\gamma_5
+ 2 g_2(Q^2)p_\mu\not\!q\gamma_5
+ \Big[g_2(Q^2)+g_3(Q^2)\vphantom{\int_0^{x_2}}\Big] q_\mu\not\!q
\gamma_5 \nnb \\
\ar \mbox{\rm other structures}\Big\} u_\Lambda(p)~, \\ \nnb \\
\label{sigmaafterb}
\Pi_{\mu}^{II}(p,q)\es
\frac{\lambda_{\Lambda_{b}}}{m_{\Lambda_{b}}^{2}-(p+q)^{2}}
\Big\{ 2 f_{1}^{T}(Q^{2}) p_\mu +
2 f_{2}^{T}(Q^{2})p_\mu\not\!q+
\Big[ f_2^{T}(Q^2) + f_3^{T}(Q^2)\Big]q_\mu\not\!q \nnb \\
\ar 2g_1^{T}(Q^2)p_{\mu}\gamma_5
- 2 g_2^{T}(Q^2)p_\mu\not\!q\gamma_5
-\Big[g_2^{T}(Q^2)+g_3^{T}(Q^2) \Big]q_\mu\not\!q\gamma_5 \nnb \\
\ar \mbox{\rm other structures} \Big\} u_\Lambda(p)~.
\eea
To compute the form factors or their combinations,
$f_{1}$, $f_{2}$, $f_{2}+f_{3}$, $g_{1}$, $g_{2}$ and
$g_{2}+g_{3}$, we will choose the independent structures $p_{\mu}$,
$p_{\mu}\rlap/q$, $q_{\mu}\rlap/q$, $p_{\mu}\gamma_5$,
$p_{\mu}\rlap/q\gamma_5$, and $q_{\mu}\rlap/q \gamma_5$ from
 Eq. (\ref{sigmaaftera}), respectively. The same structures are selected to
 calculate the form factors or their combinations labeled by T in the second
 correlation function, Eq. (\ref{sigmaafterb}).

The next step is to calculate the  correlation
functions from QCD side in deep Euclidean region where $(p+q)^2\ll0$. For this aim, we expend the time ordering products of the interpolating current of the $\Lambda_b$ and
transition currents in the correlation functions (see Eq. (\ref{T})) near the light cone,
$x^2\simeq 0$  via operator product expansion, where the short and long distance effects
are separated. The former is calculated using QCD perturbation theory, whereas the latter
are parameterized in terms of the  $\Lambda$ DA's.  Mathematically, this is equivalent to
contract out all quark pairs in the time ordering product of the $J^{\Lambda_{b}}
$ and  transition currents via the Wick's theorem. As a result of this procedure, we
obtain the following representations of the correlation functions in
QCD side:
\bea\label{mut.m}
\Pi^I_{\mu} \es \frac{-i}{\sqrt{6}} \epsilon^{abc}\int d^4x e^{iqx}
\Big\{\Big[2 ( C )_{\eta\phi} (\gamma_5)_{\rho\beta}+( C
)_{\eta\beta} (\gamma_5)_{\rho\phi}+( C
)_{\beta\phi} (\gamma_5)_{\eta\rho}\Big] \nnb \\
\ar\beta\Big[2 (C \gamma_5
)_{\eta\phi}(I)_{\rho\beta}
+ (C \gamma_5 )_{\eta\beta}(I)_{\rho\phi}+(C \gamma_5
)_{\beta\phi}(I)_{\eta\rho} \Big]\Big\} \Big[
\gamma_{\mu} (1-\gamma_5)\Big]_{\sigma\theta} \nnb \\
\cp S_b(-x)_{\beta\sigma}
\langle 0 |  u_\eta^a(0)
s_\theta^b(x) d_\phi^c(0) | \Lambda (p)\rangle~,
\eea
\bea\label{mut.mm}
\Pi^{II}_{\mu} \es \frac{-i}{\sqrt{6}} \epsilon^{abc}\int d^4x e^{iqx}
\Big\{\Big[2 ( C )_{\eta\phi} (\gamma_5)_{\rho\beta}+( C
)_{\eta\beta} (\gamma_5)_{\rho\phi}+( C
)_{\beta\phi} (\gamma_5)_{\eta\rho}\Big] \nnb \\
\ar \beta\Big[2 (C \gamma_5
)_{\eta\phi}(I)_{\rho\beta}
+ (C \gamma_5 )_{\eta\beta}(I)_{\rho\phi}+(C \gamma_5
)_{\beta\phi}(I)_{\eta\rho} \Big]\Big\} \Big[
i q^\nu \sigma_{\mu\nu} (1+\gamma_5)\Big]_{\sigma\theta} \nnb \\
\cp S_b(-x)_{\beta\sigma}
\langle 0 |  u_\eta^a(0)
s_\theta^b(x) d_\phi^c(0) | \Lambda (p)\rangle~,
\eea
The heavy quark propagator, $ S_b(x)$ is calculated in \cite{22Balitsky}:

\bea\label{heavylightguy}
 S_b (x)& =&  S_b^{free} (x) - i g_s \int \frac{d^4 k}{(2\pi)^4}
e^{-ikx} \int_0^1 dv \Bigg[\frac{\not\!k + m_Q}{( m_Q^2-k^2)^2}
G^{\mu\nu}(vx) \sigma_{\mu\nu} \nnb \\
\ar \frac{1}{m_Q^2-k^2} v x_\mu
G^{\mu\nu} \gamma_\nu \Bigg]~.
 \eea
where,
\bea\label{freeprop}
S^{free}_{b}
\es\frac{m_{b}^{2}}{4\pi^{2}}\frac{K_{1}(m_{b}\sqrt{-x^2})}{\sqrt{-x^2}}-i
\frac{m_{b}^{2}\not\!x}{4\pi^{2}x^2}K_{2}(m_{b}\sqrt{-x^2})~,
\eea
and  $K_i$ are the Bessel functions. The terms proportional to the gluon
field strength are contributed mainly to the  four and five particle
distribution functions \cite{22Balitsky,2317,2418,25Braun1b,26Liu}
and expected to be very small
in our case, hence when doing calculations,  these terms  are ignored.
The matrix element $ \epsilon^{abc}\langle 0 |  u_\eta^a(0)
d_\theta^b(0)   s_\phi^c(x) |  \Lambda (p)\rangle$  appearing in
Eqs. (\ref{mut.m},\ref{mut.mm})
represents  the $\Lambda$'s wave functions, which are calculated in
\cite{26Liu} and we list them out for the completeness of this paper in
the Appendix. Using the $\Lambda$ wave functions and the expression of
the heavy quark propagator, and after performing the Fourier transformation,
the final expressions of the correlation functions for both vertexes are
found in terms of the  $\Lambda$ DA's in QCD or theoretical side.

In order
to obtain the sum rules for the  form factors, $f_{1}$, $f_{2}$, $f_{3}$, $g_{1}$,
$g_{2}$,  $g_{3}$, $f^T_{1}$, $f^T_{2}$, $f^T_{3}$, $g^T_{1}$,
$g^T_{2}$ and $g^T_{3}$, we equate the
coefficients of the  corresponding structures from both sides of the
correlation functions through the dispersion relations and apply
Borel transformation with respect to $(p+q)^2$ to suppress the
contribution of the higher states and continuum.
The expressions for the sum rules of the form factors are very lengthy,
so we will give only extrapolation formulas to explore their dependency
on the transferred momentum squared $q^2$.

The explicit  expressions of the sum rules for the  form factors
depict that to calculate the values of the form factors, we need
also the expression of  the residue, $\lambda_{\Lambda_{b}}$. This residue is calculated in
\cite{27alievozpineci}.

Few words about the relations among the form factors in
HQET are in order. In HQET, the number of independent form factors  is reduced
to two, $F_1$ and $F_2$, so the transition matrix element can be
parameterized in terms of these two form factors as
\cite{27alievozpineci,28Mannel}:
\bea\label{matrixel1111}
\langle \Lambda(p) \mid \bar s\Gamma b\mid \Lambda_b(p+q)\rangle \es
\bar{u}_\Lambda(p)[F_1(Q^2)+\not\!vF_2(Q^2)]\Gamma u_{\Lambda_b}(p+q),
\eea
Here, $\Gamma$ refers to any Dirac matrices and
$\not\!v=({\not\!p}+{\not\!q})/m_{\Lambda_{b}}$. Comparing this
matrix element with definitions of  the form factors in Eqs.
(\ref{matrixel1a}) and (\ref{matrixel1b}),  the following
relations among the form factors are obtained (see also
\cite{29Chen,30ozpineci}):
\bea\label{matrixel22222}
 f_{1}\es g_{1}=f_{2}^{T} = g_{2}^{T}=F_1+ \frac{m_\Lambda}{m_{\Lambda_b}}
F_2~,\nnb\\
f_2 \es g_2 = f_3 = g_3=\frac{F_2}{m_{\Lambda_b}}~,\nnb\\
f_1^{T} \es g_1^{T} =\frac{F_2}{m_{\Lambda_b}}q^2~,\nnb\\
f_3^{T} \es-\frac{F_2}{m_{\Lambda_b} }(m_{\Lambda_b}-m_{\Lambda})~,\nnb\\
g_3^{T} \es\frac{F_2}{m_{\Lambda_b} }(m_{\Lambda_b}+m_{\Lambda})~.
\eea

\section{Numerical Analysis}
This section is devoted to  the numerical analysis of the form factors,
their extrapolation in terms of the momentum transferred square and
calculation of the total decay rate and branching ratio for rare
$\Lambda_{b}\rar \Lambda \ell^+\ell^- $  transition in QCD.

Some input parameters used in the numerical calculations are:
$\uu(1~GeV) = \dd(1~GeV)= -(0.243 \pm 0.01)^3~GeV^3$, $m_0^2 (1~GeV)
= (0.8\pm 0.2)~GeV^2$ \cite{R31}, $m_\Lambda = (1115.683 \pm 0.006)
~MeV$, $m_{\Lambda_{b}} =  (5620.2 \pm1.6)~ MeV$ and $m_b = (4.7\pm
0.1)~GeV$. Sum rules for the form factors depict that the $\Lambda$
DA's are the main input parameters (see the Appendix). They contain
4 independent parameters which are given as \cite{26Liu}: \bea
f_\Lambda\es(6.0\pm0.3)\times10^{-3}~ \mbox{GeV}^2~,\hspace{2.5cm}
\lambda_1=(1.0\pm0.3)\times10^{-2}~ \mbox{GeV}^2~,\nnb\\
|\lambda_2|\es(0.83\pm0.05)\times10^{-2}~
\mbox{GeV}^2~,\hspace{2.0cm}|\lambda_3|=(0.83\pm0.05)\times10^{-2}~
\mbox{GeV}^2~.\label{lambdapara}
\eea

It is well known that, the Wilson coefficient $C_9^{eff}$ receives
long distance contributions from $J/\psi$ family, in addition to
short distance contributions. In the present work, we do not take
into account the long distance effects. From the  explicit expressions
for the form factors, it is clear that they depend  on three
auxiliary parameters, continuum threshold $s_0$, Borel mass
parameter $M_B^2$ and the parameter $\beta$ entering  the most
general form of the interpolating current of the $\Lambda_b$. The
form factors should be independent of these auxiliary parameters.
Therefore, we  look for working regions for these parameters, where
the form factors are practically independent of them. To determine
the working region for the Borel mass parameter the procedure is as
follows:  the lower limit  is obtained  requiring that the higher
states and continuum contributions constitute a small percent of the
total dispersion integral. The upper limit of $M_B^2$ is chosen
demanding that  the series of the light cone expansion with
increasing twist should be convergent.  As a result, the common working
region of $M_B^2$ is found to be $15 ~GeV^{2}\leq M_{B}^{2}\leq 30~ GeV^{2}$. As an example, we present the dependence of the form factor $f_1$ 
on Borel mass parameter, $M_B^2$   at two fixed values of  $q^2$ in Fig. 1. From this figure it  follows that the form factor $f_1$ exhibits a good stability with respect to the variations of $M_B^2$.
The continuum threshold $s_{0}$ is correlated to the first exited
state with quantum numbers of the interpolating current of the
$\Lambda_b$ and  is not completely arbitrary. Numerical analysis
leads to the interval, $(m_{\Lambda_b}+0.3)^2\leq s_0\leq
(m_{\Lambda_b}+0.5)^2$, where the form factors weakly depend on the
continuum threshold.  In order to attain the working region for the  parameter, $\beta$, we look for the variation of the form
factors with respect to  $ \cos\theta$, where $\beta=tan\theta$. After performing numerical calculations, we obtained 
 that in the  interval
$-0.6\leq \cos\theta\leq 0.3$   all form factors  weakly depend on $\beta$. As an example, we  show the dependence of the form factor, $f_1$ on $ \cos\theta$  at two fixed values of the $q^2$ and at  $M_B^2=22~GeV^2$ in Fig. 2 . From this figure indeed we see that in the aforementioned region of $ \cos\theta$, the form factor $f_1$ weakly depends on $\beta$.
\begin{figure}[h!]
\caption{\label{fig1} The dependence of form factor, $f_1$ on Borel mass 
parameter  at two fixed values of the $q^2$, and at  $s_0=35~GeV^2$ and $\beta=5$.}
    \begin{center}
    \includegraphics[width=18cm]{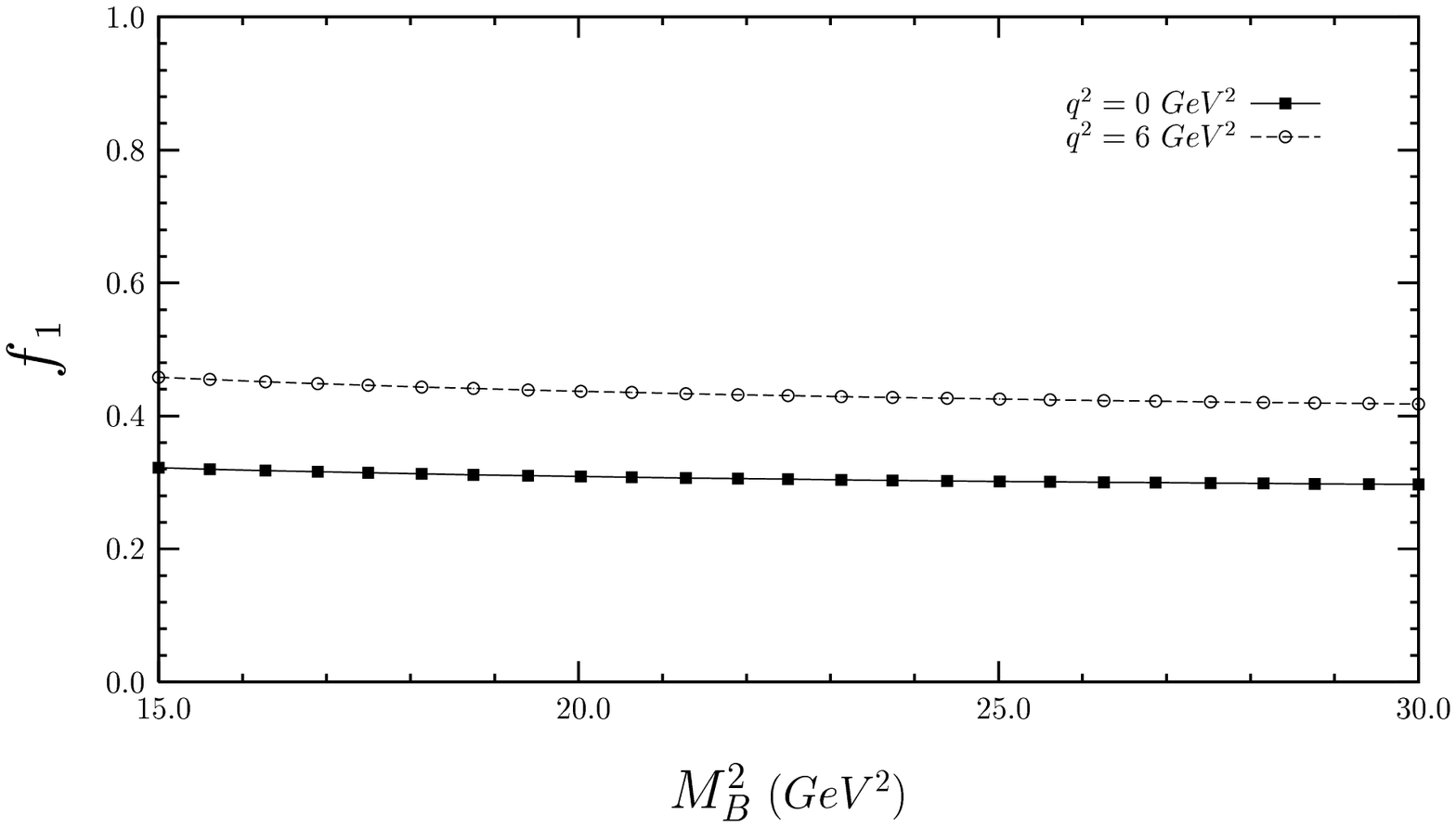}
    \end{center}
    \end{figure}
\begin{figure}[h!]
\caption{\label{fig1} The dependence of form factor, $f_1$ on $ \cos\theta$ 
parameter  at two fixed values of the $q^2$, and at $s_0=35~GeV^2$ and $M_B^2=22~GeV^2$.}
    \begin{center}
    \includegraphics[width=18cm]{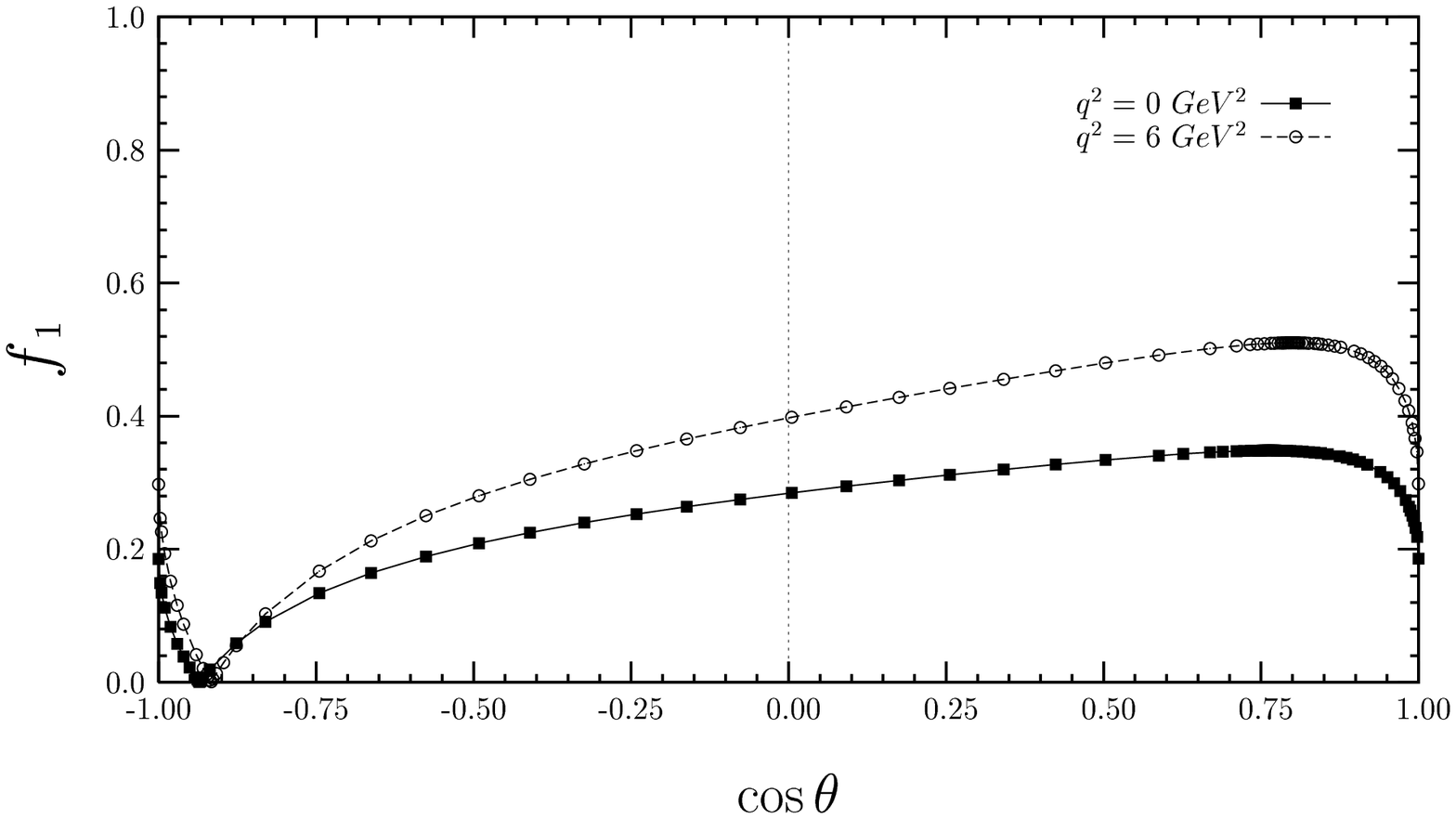}
    \end{center}
    \end{figure}

The analysis of the sum rules, as has already been explained above,
is based on, so called, the standard procedure, i.e., the
continuum threshold $s_0$ is independent of $M_B^2$ and $q^2$.  However, in \cite{R32}, instead of the  standard procedure, namely, independence of the $s_0$
 from  $M_B^2$ and $q^2$, it is assumed that the continuum threshold  depends on $M_B^2$ and $q^2$ and  this leads to large   realistic errors. Following \cite{R32},
in the present work the systematic error is  taken to be around
15\%.
\begin{table}[h]
\renewcommand{\arraystretch}{1.5}
\addtolength{\arraycolsep}{3pt}
$$
\begin{array}{|c|c|c|c|}
\hline \hline
     & \multicolumn{3}{c|}{\mbox{QCD sum rules}} \\
\hline
                & \mbox{a}            & \mbox{b}  & m_{fit}^2   \\
\hline
 f_1            &  -0.046  &   0.368  &  39.10     \\
 f_2            &   0.0046  &  -0.017  &  26.37     \\
 f_3            &   0.006  &  -0.021  &  22.99     \\
 g_1            &  -0.220  &   0.538  &  48.70     \\
 g_2            &   0.005  &  -0.018  &  26.93     \\
 g_3            &   0.035  &  -0.050  &  24.26     \\
 f_2^{T}        &  -0.131  &   0.426  &  45.70     \\
 f_3^{T}        &  -0.046  &   0.102  &  28.31     \\
 g_2^{T}        &  -0.369  &   0.664  &  59.37     \\
 g_3^{T}        &  -0.026  &  -0.075  &  23.73     \\
\hline \hline
\end{array}
$$
\caption{Parameters appearing in  the fit function of the  form
factors, $f_{1}$, $f_{2}$, $f_{3}$, $g_{1}$,
$g_{2}$,  $g_{3}$,  $f^T_{2}$, $f^T_{3}$,
$g^T_{2}$ and $g^T_{3}$  in full theory for $\Lambda_{b}\rightarrow \Lambda
\ell^{+}\ell^{-}$. In this Table only central values of the parameters are presented.} \label{tab:13}
\renewcommand{\arraystretch}{1}
\addtolength{\arraycolsep}{-1.0pt}
\end{table}
\begin{table}[h]
\renewcommand{\arraystretch}{1.5}
\addtolength{\arraycolsep}{3pt}

$$
\begin{array}{|c|c|c|c|}
\hline \hline
     & \multicolumn{3}{c|}{\mbox{QCD sum rules}} \\
\hline
     & \mbox{c} & m_{fit}^{' 2} & m_{fit}^{'' 2}      \\
\hline
 f_1^{T}  &   -1.191  &  23.81   &  59.96      \\
 g_1^{T}  &   -0.653  &  24.15   &  48.52      \\
 \hline \hline
\end{array}
$$
\caption{Parameters appearing in  the fit function of the  form
factors $f^T_{1}$ and $g^T_{1}$ in full theory for $\Lambda_{b}\rightarrow \Lambda
\ell^{+}\ell^{-}$.} \label{tab:133}
\renewcommand{\arraystretch}{1}
\addtolength{\arraycolsep}{-1.0pt}
\end{table}

In calculating the branching ratio of the $\Lambda_b \rar \Lambda
\ell^+ \ell^-$ decay, the dependence of the form factors $f_i(q^2)$,
$g_i(q^2)$, $f_i^T(q^2)$, and $g_i^T(q^2)$ on $q^2$ in the physical
region $4 m_\ell^2 \le q^2 \le (m_{\Lambda_b}-m_\Lambda)^2$ are
needed. But unfortunately, sum rules predictions for the form
factors are not reliable in the whole physical region. Therefore, in
order to obtain the $q^2$ dependence of the form factors from sum
rules we consider a range of $q^2$ where the correlation function
can reliably be calculated. For this aim we choose a region which is
approximately $1~GeV$ below the perturbative cut, i.e., up to $q^2
\simeq 12~GeV^2$. To be able to extend the results for the form
factors to the whole physical region, we look for a parameterization
of the form factors in such a way that, in the region $4 m_\ell^2
\le q^2 \le 12~GeV^2$ this parameterization coincides with the sum
rules predictions.

The next step is to present the $q^2$ dependency of the form
factors. Our numerical calculations show that the best
parameterization for the dependence of the form factors $f_{1}$,
$f_{2}$, $f_{3}$, $g_{1}$, $g_{2}$,  $g_{3}$,  $f^T_{2}$, $f^T_{3}$,
$g^T_{2}$ and $g^T_{3}$ on $q^2$ is as follows: \bea\label{17au}
 f_{i}(q^2)[g_{i}(q^2)]=\frac{a}{\Bigg(1-\ds\frac{q^2}{m_{fit}^2}\Bigg)}+
\frac{b}{\Bigg(1-\ds\frac{q^2}{m_{fit}^2}\Bigg)^2}~,
\eea
 where the fit parameters
$a,~b$ and $m_{fit}^2$ in full theory   are given in Table
\ref{tab:13}. On the other hand, we find that the best fit for the form factors
$f^T_{1}$ and $g^T_{1}$ is of the following form, \bea\label{1777au}
 f^T_{1}(q^2)[g^T_{1}(q^2)]=\frac{c}{\Bigg(1-\ds\frac{q^2}{m_{fit}^{' 2}}\Bigg)}-
\frac{c}{\Bigg(1-\ds\frac{q^2}{m_{fit}^{'' 2}}\Bigg)^2}~. \eea The
results for the parameters $c$, $m^{'2}_{fit}$ and $m^{''2}_{fit}$
are presented in Table \ref{tab:133}. In extraction of the values of the fit parameters presented in both Tables \ref{tab:13} and  \ref{tab:133},  the values of the continuum threshold, $s_0=35~GeV^2$, Borel mass parameter, $M_B^2=22~GeV^2$ and  $\cos\theta=0.2$ have been used.

The values of form factors at $q^2=0$  are also presented in Table
\ref{tab:15}. In this table we also present the numerical results
obtained from HQET, using the values for the form factors
$F_1(0)=0.462$ and $F_2 = -0.077$ predicted in \cite{yenison}, and
relations in Eq. (\ref{matrixel22222}) at HQET limit. The errors in
the values of the form factors at $q^2=0$ are due to the
uncertainties coming from $M_B^2$, $s_0$, the  parameter $\beta$,
errors in the input parameters, as well as from the systematic
errors. From this Table we see that, the predictions of the HQET on the form factors
are changed more than 40\% for the form factors $f_1(0)$,
$g_1(0)$, $f_2^T(0)$, and $g_1^T(0)$, while the results of both approaches
are very close to each other for the remaining form factors.

\begin{table}[h]
\renewcommand{\arraystretch}{1.5}
\addtolength{\arraycolsep}{3pt}

$$
\begin{array}{|c|c|c|}                                  \hline \hline
 &\mbox{Present work} & \mbox{HQET (\cite{13Huang}) }  \\
                            \hline
 f_{1}(0)     &  0.322 \pm 0.112  &  0.446        \\
 f_{2}(0)     & -0.011 \pm 0.004  & -0.013        \\
 f_{3}(0)     & -0.015 \pm 0.005  & -0.013        \\
 g_{1}(0)     &  0.318 \pm 0.110  &  0.446        \\
 g_{2}(0)     & -0.013 \pm 0.004  & -0.013        \\
 g_{3}(0)     & -0.014 \pm 0.005  & -0.013        \\
 f_{1}^{T}(0) &  0     \pm 0.0    &  0.0          \\
 f_{2}^{T}(0) &  0.295 \pm 0.105  &  0.446        \\
 f_{3}^{T}(0) &  0.056 \pm 0.018  & 0.061        \\
 g_{1}^{T}(0) &  0     \pm 0.0    &  0.0          \\
 g_{2}^{T}(0) &  0.294 \pm 0.105  &  0.446        \\
 g_{3}^{T}(0) & -0.101 \pm 0.035  & -0.092        \\
 \hline \hline
\end{array}
$$

\caption{The values of the form factors at $q^2=0$  for
$\Lambda_{b}\rightarrow \Lambda \ell^{+}\ell^{-}$.} \label{tab:15}
\renewcommand{\arraystretch}{1}
\addtolength{\arraycolsep}{-1.0pt}
\end{table}

The final  task is to calculate the total decay rate of the
$\Lambda_{b}\rar \Lambda \ell^+\ell^-$ transition in the whole
physical region,  $ 4m_\ell^2 \leq q^2 \leq (m_{\Lambda_{b}} -
m_{\Lambda})^2$.  The differential decay rate is obtained as: \bea
\frac{d\Gamma}{ds} = \frac{G^2\alpha^2_{em} m_{\Lambda_b}}{8192
\pi^5}| V_{tb}V_{ts}^*|^2 v \sqrt{\lambda} \, \Bigg[
\Theta(s) + \frac{1}{3} \Delta(s)\Bigg]~,  \label{rate} \eea where
$s= q^2/m^2_{\Lambda_b}$, $r= m^2_{\Lambda}/m^2_{\Lambda_b}$, $\lambda=\lambda(1,r,s)=1+r^2+s^2-2r-2s-2rs$,
$G_F = 1.17 \times 10^{-5}$ GeV$^{-2}$ is the Fermi coupling
constant and  $v=\sqrt{1-\frac{4 m_\ell^2}{q^2}}$ is the
lepton velocity. For the element of the CKM matrix $\mid
V_{tb}V_{ts}^\ast\mid=0.041$ has been used~\cite{33Yao:2006px}. The
functions $\Theta(s)$ and $\Delta(s)$ are given as:

\bea
\Theta(s) \es
32 m_\ell^2 m_{\Lambda_b}^4 s (1+r-s) \ga \vel D_3 \ver^2 +
\vel E_3 \ver^2 \dr \nnb \\
\ar 64 m_\ell^2 m_{\Lambda_b}^3 (1-r-s) \, \mbox{\rm Re} [D_1^\ast E_3 + D_3
E_1^\ast] \nnb \\
\ar 64 m_{\Lambda_b}^2 \sqrt{r} (6 m_\ell^2 - m_{\Lambda_b}^2 s)
{\rm Re} [D_1^\ast E_1] \nnb \\
\ar 64 m_\ell^2 m_{\Lambda_b}^3 \sqrt{r}
\Big( 2 m_{\Lambda_b} s {\rm Re} [D_3^\ast E_3] + (1 - r + s)
{\rm Re} [D_1^\ast D_3 + E_1^\ast E_3]\Big) \nnb \\
\ar 32 m_{\Lambda_b}^2 (2 m_\ell^2 + m_{\Lambda_b}^2 s)
\Big\{ (1 - r + s) m_{\Lambda_b} \sqrt{r} \,
\mbox{\rm Re} [A_1^\ast A_2 + B_1^\ast B_2] \nnb \\
\ek m_{\Lambda_b} (1 - r - s) \, \mbox{\rm Re} [A_1^\ast B_2 + A_2^\ast B_1] -
2 \sqrt{r} \Big( \mbox{\rm Re} [A_1^\ast B_1] + m_{\Lambda_b}^2 s \,
\mbox{\rm Re} [A_2^\ast B_2] \Big) \Big\} \nnb \\
\ar 8 m_{\Lambda_b}^2 \Big\{ 4 m_\ell^2 (1 + r - s) +
m_{\Lambda_b}^2 \Big[(1-r)^2 - s^2 \Big]
\Big\} \ga \vel A_1 \ver^2 +  \vel B_1 \ver^2 \dr \nnb \\
\ar 8 m_{\Lambda_b}^4 \Big\{ 4 m_\ell^2 \Big[ \lambda +
(1 + r - s) s \Big] +
m_{\Lambda_b}^2 s \Big[(1-r)^2 - s^2 \Big]
\Big\} \ga \vel A_2 \ver^2 +  \vel B_2 \ver^2 \dr \nnb \\
\ek 8 m_{\Lambda_b}^2 \Big\{ 4 m_\ell^2 (1 + r - s) -
m_{\Lambda_b}^2 \Big[(1-r)^2 - s^2 \Big]
\Big\} \ga \vel D_1 \ver^2 +  \vel E_1 \ver^2 \dr \nnb \\
\ar 8 m_{\Lambda_b}^5 s v^2 \Big\{
- 8 m_{\Lambda_b} s \sqrt{r}\, \mbox{\rm Re} [D_2^\ast E_2] +
4 (1 - r + s) \sqrt{r} \, \mbox{\rm Re}[D_1^\ast D_2+E_1^\ast E_2]\nnb \\
\ek 4 (1 - r - s) \, \mbox{\rm Re}[D_1^\ast E_2+D_2^\ast E_1] +
m_{\Lambda_b} \Big[(1-r)^2 -s^2\Big]
\ga \vel D_2 \ver^2 + \vel E_2 \ver^2\dr \Big\}~,
\eea
\bea
\Delta \left( s\right) \es
- 8 m_{\Lambda_b}^4 v^2 \lambda
\ga \vel A_1 \ver^2 + \vel B_1 \ver^2 + \vel D_1 \ver^2
+ \vel E_1 \ver^2 \dr \nnb \\
\ar 8 m_{\Lambda_b}^6 s v^2 \lambda
\Big( \vel A_2 \ver^2 +
\vel B_2 \ver^2 + \vel D_2 \ver^2 + \vel E_2 \ver^2  \Big)~,
\eea
where
\bea
\label{a9}
A_1 \es \frac{1}{q^2}\ga
f_1^T+g_1^T \dr \ga -2 m_b C_7\dr + \ga f_1-g_1 \dr C_9^{eff} \nnb \\
A_2 \es A_1 \ga 1 \rar 2 \dr ~,\nnb \\
A_3 \es A_1 \ga 1 \rar 3 \dr ~,\nnb \\
B_1 \es A_1 \ga g_1 \rar - g_1;~g_1^T \rar - g_1^T \dr ~,\nnb \\
B_2 \es B_1 \ga 1 \rar 2 \dr ~,\nnb \\
B_3 \es B_1 \ga 1 \rar 3 \dr ~,\nnb \\
D_1 \es \ga f_1-g_1 \dr C_{10} ~,\nnb \\
D_2 \es D_1 \ga 1 \rar 2 \dr ~, \\
D_3 \es D_1 \ga 1 \rar 3 \dr ~,\nnb \\
E_1 \es D_1 \ga g_1 \rar - g_1 \dr ~,\nnb \\
E_2 \es E_1 \ga 1 \rar 2 \dr ~,\nnb \\
E_3 \es E_1 \ga 1 \rar 3 \dr ~.
\eea
Integrating  the differential decay rate on $s$ in the whole physical region
$4 m_\ell^2/m_{\Lambda_b}^2 \le s \le (1-\sqrt{r})^2$ and
using the life time of the $\Lambda_b$ baryon, $\tau_{\Lambda_b}=1.383\times
10^{-12}~s$ \cite{33Yao:2006px}, we obtain the results for the branching
ratio which are
presented in  Table \ref{tab:27}.

\begin{table}[h] \centering
\begin{tabular}{|l||c|c|} \hline &
Present work  & HQET(\cite{18chen2}) \\ \hline\hline
$Br(\Lambda_{b} \rar \Lambda e^+e^-)$ & $(4.6 \pm 1.6) \times 10^{-6}$      & $(2.23 \div 3.34)   \times 10^{-6}$
\\ \hline
$Br(\Lambda_{b} \rar \Lambda \mu^+\mu^-)$ & $(4.0 \pm 1.2) \times 10^{-6}$  & $(2.08 \div 3.19)   \times 10^{-6}$
\\ \hline
$Br(\Lambda_{b}\rar \Lambda \tau^+\tau^-)$ & $(0.8 \pm 0.3) \times 10^{-6}$ & $(0.179 \div 0.276) \times 10^{-6}$
\\ \hline
\end{tabular}
\vspace{0.8cm} \caption{Values of the Branching ratio for
$\Lambda_{b}\rar \Lambda~\ell^+\ell^-$ in full theory and HQET for
different leptons .} \label{tab:27}
\end{table}

In this Table we also present the values of the branching ratio
obtained in HQET \cite{18chen2}. Comparing the results of both
approaches, we see that our predictions on the branching ratios for
the $\Lambda_b \rar \Lambda e^+ e^-$, $\Lambda_b \rar \Lambda \mu^+
\mu^-$ channels are larger approximately by a factor of two than the
ones predicted by the HQET, while for the $\Lambda_b \rar \Lambda
\tau^+ \tau^-$ channel our prediction is four times larger than the
result of the HQET. Since $10^{10} \div 10^{11}$ pairs are expected
to be produced per year at LHCb \cite{R34}, the results presented in
Table--4 show that detectability of $\Lambda_b \rar \Lambda \ell^+
\ell^-~(\ell = e,\mu,\tau)$ decays in this machine is quite high.

In conclusion, we calculate all twelve form factors responsible for
the $\Lambda_b \rar \Lambda \ell^+ \ell^-$ decay within light cone
sum rules. It is obtained  the maximum difference between our
results and HQET predictions on the form factors is about 40\%.
Using the parametrization for the form factors, the branching ratio
of the $\Lambda_b \rar \Lambda \ell^+ \ell^-$ decay is estimated,
and the result we obtain allows us to conclude that the
delectability of this decay at LHCb is quite high.

\newpage

\section*{Appendix }

In this Appendix, the general decomposition of the matrix element,
$ \epsilon^{abc}\langle 0 |  u_\eta^a(0) s_\theta^b(x) d_\phi^c(0)
|  \Lambda (p)\rangle$  entering  Eqs. (\ref{mut.m},\ref{mut.mm})
 as well as the $\Lambda$ DA's are
given \cite{26Liu}: \bea\label{wave func}
&&4\langle0|\epsilon^{abc}u_\alpha^a(a_1 x)s_\beta^b(a_2
x)d_\gamma^c(a_3 x)|\Lambda(p)\rangle\nnb\\
\es\mathcal{S}_1m_{\Lambda}C_{\alpha\beta}(\gamma_5\Lambda)_{\gamma}+
\mathcal{S}_2m_{\Lambda}^2C_{\alpha\beta}(\rlap/x\gamma_5\Lambda)_{\gamma}\nnb\\
\ar
\mathcal{P}_1m_{\Lambda}(\gamma_5C)_{\alpha\beta}\Lambda_{\gamma}+
\mathcal{P}_2m_{\Lambda}^2(\gamma_5C)_{\alpha\beta}(\rlap/x\Lambda)_{\gamma}+
(\mathcal{V}_1+\frac{x^2m_{\Lambda}^2}{4}\mathcal{V}_1^M)(\rlap/pC)_{\alpha\beta}(\gamma_5\Lambda)_{\gamma}
\nnb\\\ar
\mathcal{V}_2m_{\Lambda}(\rlap/pC)_{\alpha\beta}(\rlap/x\gamma_5\Lambda)_{\gamma}+
\mathcal{V}_3m_{\Lambda}(\gamma_\mu
C)_{\alpha\beta}(\gamma^\mu\gamma_5\Lambda)_{\gamma}+
\mathcal{V}_4m_{\Lambda}^2(\rlap/xC)_{\alpha\beta}(\gamma_5\Lambda)_{\gamma}\nnb\\\ar
\mathcal{V}_5m_{\Lambda}^2(\gamma_\mu
C)_{\alpha\beta}(i\sigma^{\mu\nu}x_\nu\gamma_5\Lambda)_{\gamma} +
\mathcal{V}_6m_{\Lambda}^3(\rlap/xC)_{\alpha\beta}(\rlap/x\gamma_5\Lambda)_{\gamma}
+(\mathcal{A}_1\nnb\\
\ar\frac{x^2m_{\Lambda}^2}{4}\mathcal{A}_1^M)(\rlap/p\gamma_5
C)_{\alpha\beta}\Lambda_{\gamma}+
\mathcal{A}_2m_{\Lambda}(\rlap/p\gamma_5C)_{\alpha\beta}(\rlap/x\Lambda)_{\gamma}+
\mathcal{A}_3m_{\Lambda}(\gamma_\mu\gamma_5
C)_{\alpha\beta}(\gamma^\mu \Lambda)_{\gamma}\nnb\\\ar
\mathcal{A}_4m_{\Lambda}^2(\rlap/x\gamma_5C)_{\alpha\beta}\Lambda_{\gamma}+
\mathcal{A}_5m_{\Lambda}^2(\gamma_\mu\gamma_5
C)_{\alpha\beta}(i\sigma^{\mu\nu}x_\nu \Lambda)_{\gamma}+
\mathcal{A}_6m_{\Lambda}^3(\rlap/x\gamma_5C)_{\alpha\beta}(\rlap/x
\Lambda)_{\gamma}\nnb\\\ar(\mathcal{T}_1+\frac{x^2m_{\Lambda}^2}{4}\mathcal{T}_1^M)(p^\nu
i\sigma_{\mu\nu}C)_{\alpha\beta}(\gamma^\mu\gamma_5
\Lambda)_{\gamma}+\mathcal{T}_2m_{\Lambda}(x^\mu p^\nu
i\sigma_{\mu\nu}C)_{\alpha\beta}(\gamma_5
\Lambda)_{\gamma}\nnb\\\ar
\mathcal{T}_3m_{\Lambda}(\sigma_{\mu\nu}C)_{\alpha\beta}(\sigma^{\mu\nu}\gamma_5
\Lambda)_{\gamma}+
\mathcal{T}_4m_{\Lambda}(p^\nu\sigma_{\mu\nu}C)_{\alpha\beta}(\sigma^{\mu\rho}x_\rho\gamma_5
\Lambda)_{\gamma}\nnb\\\ar \mathcal{T}_5m_{\Lambda}^2(x^\nu
i\sigma_{\mu\nu}C)_{\alpha\beta}(\gamma^\mu\gamma_5
\Lambda)_{\gamma}+ \mathcal{T}_6m_{\Lambda}^2(x^\mu p^\nu
i\sigma_{\mu\nu}C)_{\alpha\beta}(\rlap/x\gamma_5
\Lambda)_{\gamma}\nnb\\
\ar
\mathcal{T}_7m_{\Lambda}^2(\sigma_{\mu\nu}C)_{\alpha\beta}(\sigma^{\mu\nu}\rlap/x\gamma_5
\Lambda)_{\gamma}+
\mathcal{T}_8m_{\Lambda}^3(x^\nu\sigma_{\mu\nu}C)_{\alpha\beta}(\sigma^{\mu\rho}x_\rho\gamma_5
\Lambda)_{\gamma}~.\nnb~~~~~~~~~~~~~~~~~~~~~~~(A.1) \eea The
calligraphic functions in the above expression  have not definite
twists but they can be written in terms of the Lambda distribution
amplitudes (DA's) with definite and  increasing twists via   the
scalar product $px$ and the parameters $a_i$, $i=1,2,3$. The
explicit expressions for  scalar, pseudo-scalar, vector, axial
vector and tensor DA's for Lambda are given in Tables
\ref{tab:1}, \ref{tab:2}, \ref{tab:3}, \ref{tab:4} and
\ref{tab:5}, respectively.
\begin{table}[h]
\centering
\begin{tabular}{|c|} \hline
$\mathcal{S}_1 = S_1$\\ \hline\hline
 $2px\mathcal{S}_2=S_1-S_2$ \\ \hline
   \end{tabular}
\vspace{0.3cm} \caption{Relations between the calligraphic functions
and Lambda scalar DA's.}\label{tab:1}
\end{table}
\begin{table}[h]
\centering
\begin{tabular}{|c|} \hline
  $\mathcal{P}_1=P_1$\\ \hline
  $2px\mathcal{P}_2=P_1-P_2$ \\ \hline
   \end{tabular}
\vspace{0.3cm} \caption{Relations between the calligraphic functions
and Lambda pseudo-scalar DA's.}\label{tab:2}
\end{table}
\begin{table}[h]
\centering
\begin{tabular}{|c|} \hline
  $\mathcal{V}_1=V_1$ \\ \hline
  $2px\mathcal{V}_2=V_1-V_2-V_3$ \\ \hline
  $2\mathcal{V}_3=V_3$ \\ \hline
  $4px\mathcal{V}_4=-2V_1+V_3+V_4+2V_5$ \\ \hline
  $4px\mathcal{V}_5=V_4-V_3$ \\ \hline
  $4(px)^2\mathcal{V}_6=-V_1+V_2+V_3+V_4
 + V_5-V_6$ \\ \hline
 \end{tabular}
\vspace{0.3cm} \caption{Relations between the calligraphic functions
and Lambda vector DA's.}\label{tab:3}
\end{table}
\begin{table}[h]
\centering
\begin{tabular}{|c|} \hline
  $\mathcal{A}_1=A_1$ \\ \hline
  $2px\mathcal{A}_2=-A_1+A_2-A_3$ \\ \hline
   $2\mathcal{A}_3=A_3$ \\ \hline
  $4px\mathcal{A}_4=-2A_1-A_3-A_4+2A_5$ \\ \hline
  $4px\mathcal{A}_5=A_3-A_4$ \\ \hline
  $4(px)^2\mathcal{A}_6=A_1-A_2+A_3+A_4-A_5+A_6$ \\ \hline
 \end{tabular}
\vspace{0.3cm} \caption{Relations between the calligraphic functions
and Lambda axial vector DA's.}\label{tab:4}
\end{table}
\begin{table}[h]
\centering
\begin{tabular}{|c|} \hline
  $\mathcal{T}_1=T_1$ \\ \hline
  $2px\mathcal{T}_2=T_1+T_2-2T_3$ \\ \hline
  $2\mathcal{T}_3=T_7$ \\ \hline
  $2px\mathcal{T}_4=T_1-T_2-2T_7$ \\ \hline
  $2px\mathcal{T}_5=-T_1+T_5+2T_8$ \\ \hline
  $4(px)^2\mathcal{T}_6=2T_2-2T_3-2T_4+2T_5+2T_7+2T_8$ \\ \hline
  $4px \mathcal{T}_7=T_7-T_8$\\ \hline
  $4(px)^2\mathcal{T}_8=-T_1+T_2 +T_5-T_6+2T_7+2T_8$\\ \hline
 \end{tabular}
\vspace{0.3cm} \caption{Relations between the calligraphic functions
and Lambda tensor DA's.}\label{tab:5}
\end{table}

Every distribution amplitude $F(a_ipx)$=  $S_i$,
$P_i$, $V_i$, $A_i$, $T_i$ can be represented as:
\bea\label{dependent1}
F(a_ipx)=\int dx_1dx_2dx_3\delta(x_1+x_2+x_3-1) e^{ip
x\Sigma_ix_ia_i}F(x_i)~.\nnb~~~~~~~~~~~~~~~~~~~~~~~~~~~~~~(A.2)
\eea
where, $x_{i}$ with $i=1,~2$ and $3$ are longitudinal momentum
fractions carried by the participating quarks.

The explicit expressions for the $\Lambda$ DA's up to twist 6 are given as:
twist-$3$ DA's:
\bea
V_1(x_i)\es0~,\hspace{4.5cm}A_1(x_i)=-120x_1x_2x_3\phi_3^0~,\nnb\\
T_1(x_i)\es0~.\nnb~~~~~~~~~~~~~~~~~~~~~~~~~~~~~~~~~~~~~~~~~~~~~~~~~~~~~~~~~~~~~~~~~~~~~~~~~~~~~~~~~~~~(A.3)
\eea
twist-$4$ DA's:
\bea
S_1(x_i)\es6x_3(1-x_3)(\xi_4^0+\xi_4^{'0})~,\hspace{1.5cm}P_1(x_i)=6(1-x_3)(\xi_4^0-\xi_4^{'0})~,\nnb\\
V_2(x_i)\es0~,\hspace{5.0cm}A_2(x_i)=-24x_1x_2\phi_4^0~,\nnb\\
V_3(x_i)\es12(x_1-x_2)x_3\psi_4^0~,\hspace{2.3cm}A_3(x_i)=-12x_3(1-x_3)\psi_4^0~,\nnb\\
T_2(x_i)\es0~,\hspace{5.0cm}T_3(x_i)=6(x_2-x_1)x_3(-\xi_4^0+\xi_4^{'0})~,\nnb\\
T_7(x_i)\es-6(x_1-x_2)x_3(\xi_4^0+\xi_4^{'0})~.\nnb~~~~~~~~~~~~~~~~~~~~~~~~~~~~~~~~~~~~~~~~~~~~~~~~~~~~~~(A.4)
\eea
Ttwist-$5$ DA's:
\bea
S_2(x_i)\es\frac32(x_1+x_2)(\xi_5^0+\xi_5^{'0})~,\hspace{1.5cm}P_2(x_i)=\frac32(x_1+x_2)(\xi_5^0-\xi_5^{'0})~,\nnb\\
V_4(x_i)\es3(x_2-x_1)\psi_5^0~,\hspace{2.8cm}A_4(x_i)=-3(1-x_3)\psi_5^0~,\nnb\\
V_5(x_i)\es0~,\hspace{4.9cm}A_5(x_i)=-6x_3\phi_5^0~,\nnb\\
T_4(x_i)\es-\frac32(x_1-x_2)(\xi_5^0+\xi_5^{'0})~,\hspace{1.3cm}T_5(x_i)=0~,\nnb\\
T_8(x_i)\es-\frac32(x_1-x_2)(\xi_5^0-\xi_5^{'0})~.\nnb~~~~~~~~~~~~~~~~~~~~~~~~~~~~~~~~~~~~~~~~~~~~~~~~~~~~~~~(A.5)
\eea
and twist-$6$ DA's:
\bea
V_6(x_i)\es0~,\hspace{4.5cm}A_6(x_i)=-2\phi_6^0~,\nnb\\
T_6(x_i)\es0~.\nnb~~~~~~~~~~~~~~~~~~~~~~~~~~~~~~~~~~~~~~~~~~~~~~~~~~~~~~~~~~~~~~~~~~~~~~~~~~~~~~~~~~(A.6)
\eea
 The following functions are encountered to the above amplitudes and they can be  defined in
terms of the  4 independent parameters, namely  $f_\Lambda$,
$\lambda_1$, $\lambda_2$ and $\lambda_3$:
\bea
\phi_3^0\es\phi_6^0=-f_\Lambda~,\hspace{2.8cm}\phi_4^0=\phi_5^0=-\frac12(f_\Lambda+\lambda_1)~,\nnb\\
\psi_4^0\es\psi_5^0=\frac12(f_\Lambda-\lambda_1)~,\hspace{1.6cm}\xi_4^0=\xi_5^0=\lambda_2+\lambda_3~,\nnb\\
\xi_4^{'0}\es\xi_5^{'0}=\lambda_3-\lambda_2~.\nnb~~~~~~~~~~~~~~~~~~~~~~~~~~~~~~~~~~~~~~~~~~~~~~~~~~~~~~~~~~~~~~~~~~~~(A.7)
\eea
\end{document}